\documentclass[Physsubmission, Phys]{SciPost}

\hypersetup{
    colorlinks,
    linkcolor={red!50!black},
    citecolor={blue!50!black},
    urlcolor={blue!80!black}
}

\usepackage[bitstream-charter]{mathdesign}
\newcommand\be{\begin{equation}}
\newcommand\ee{\end{equation}}
\newcommand\bea{\begin{eqnarray}}
\newcommand\eea{\end{eqnarray}}

\newcommand\dd{{\rm d}}

 \def\mf {\mathfrak}
 \DeclareMathOperator\spn{span}
\def\L{\Lambda}

 \def\>#1{{\bf #1}} 
 
  \newcommand\ro{\eta }
  \newcommand\esf{{\mathcal S}}

\usepackage{amsthm}
\newtheorem{proposition}{Proposition}

\newcommand\w{\wedge}

  \newcommand\til{{\rm tl}}

\DeclareSymbolFont{usualmathcal}{OMS}{cmsy}{m}{n}
\DeclareSymbolFontAlphabet{\mathcal}{usualmathcal}

\begin{document}

\begin{center}{\Large \textbf{
A general approach to noncommutative spaces from   Poisson homogeneous spaces: Applications to (A)dS and Poincar\'e\\
}}\end{center}

\begin{center}
Angel Ballesteros\textsuperscript{1},
Ivan Gutierrez-Sagredo\textsuperscript{2$\star$} and
Francisco J.~Herranz\textsuperscript{1$\star$}
\end{center}

\begin{center}
{\bf 1} Departamento de F\'isica, Universidad de Burgos, 
09001 Burgos, Spain
\\
{\bf 2} Departamento de Matem\'aticas\&Computaci\'on, Universidad de Burgos, 
09001 Burgos, Spain
\\
 {\href{mailto:igsagredo@ubu.es}{igsagredo@ubu.es}, \href{mailto:fjherranz@ubu.es}{fjherranz@ubu.es}}
\end{center}

\begin{center}
\today
\end{center}


\definecolor{palegray}{gray}{0.95}
\begin{center}
\colorbox{palegray}{
  \begin{tabular}{rr}
  \begin{minipage}{0.1\textwidth}
    \includegraphics[width=20mm]{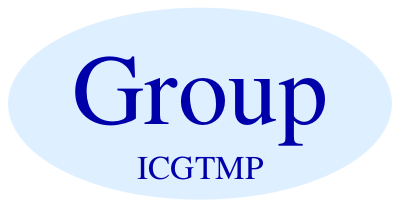}
  \end{minipage}
  &
  \begin{minipage}{0.85\textwidth}
    \begin{center}
    {\it 34th International Colloquium on Group Theoretical Methods in Physics}\\
    {\it Strasbourg, 18-22 July 2022} \\
    \doi{10.21468/SciPostPhysProc.?}\\
    \end{center}
  \end{minipage}
\end{tabular}
}
\end{center}

\section*{Abstract}
{\bf
In this contribution we present a general procedure that allows the construction of noncommutative spaces with quantum group invariance as the quantization of their associated coisotropic Poisson homogeneous spaces coming from   a coboundary Lie bialgebra structure. The approach is illustrated by obtaining in an explicit form several noncommutative spaces from   (3+1)D (A)dS and Poincar\'e coisotropic Lie bialgebras. In particular, we review the construction of the $\kappa$-Minkowski and $\kappa$-(A)dS spacetimes in terms of the cosmological constant $\L$. Furthermore, we present all noncommutative Minkowski and (A)dS spacetimes that preserved a quantum Lorentz subgroup. Finally, it is also shown that the same setting can be used to construct the three possible  6D $\kappa$-Poincar\'e   spaces of time-like worldlines. Some open problems are also addressed.
}

\vspace{10pt}
\noindent\rule{\textwidth}{1pt}
\tableofcontents\thispagestyle{fancy}
\noindent\rule{\textwidth}{1pt}
\vspace{10pt}


\section{Introduction}
 \label{s1}

The aim of this contribution is twofold. Firstly, we present a systematic "six-step" procedure that allows the construction of   different noncommutative spaces with a common underlying homogeneous space $G/H$ where $G$ is a Lie group and $H$ is  the isotropy Lie subgroup. The approach requires starting with a coboundary Lie bialgebra $(\mf g,\delta(r))$ such that $\mf g$ is the Lie algebra of $G$ and $\delta$ is the cocommutator obtained from a classical $r$-matrix $r$~\cite{ChariPressley1994,Drinfeld1983hamiltonian}. The main requirement for our development is that $\delta$ must satisfy the coisotropic condition $\delta(\mathfrak h) \subset \mathfrak h \wedge \mathfrak g$ with respect to the isotropy Lie algebra $\mf h$ of $H$~\cite{Ciccoli2006, BGM2019coreductive,GH2021symmetry}. Since 
 coboundary Lie bialgebras  are the tangent counterpart of    Poisson--Lie  groups   $(G,\Pi)$ with a Poisson structure $\Pi$, the latter just comes from the so-called Sklyanin bracket in this quantum group setting. Therefore,  this leads to 
coisotropic Poisson homogeneous spaces  $(G/H,\pi)$ where  the Poisson structure $\pi$  on $G/H$ is obtained  via canonical projection of the Poisson--Lie structure $\Pi$ on the Lie group $G$. The quantization of $(G/H,\pi)$ gives rise to the corresponding noncommutative space.

Secondly, we illustrate this approach by reviewing, from this general perspective,    several  very recent noncommutative spaces  that could be   of      interest    in a quantum gravity framework~\cite{Addazi:2021xuf}. In particular, throughout the paper  we will   focus on the (3+1)D  (Anti-)de Sitter (in short (A)dS)) and Poincar\'e Lie groups   and  their associated  (3+1)D homogeneous spacetimes together with  the 6D Poincar\'e homogeneous space of time-like geodesics.

 The structure of the paper is as follows. In the next section we recall the main necessary mathematical notions and geometric structures.   And, as the main result, we present the six-step approach to noncommutative spaces from 
 coisotropic Poisson homogeneous spaces. In Section~\ref{s3} we apply this procedure   in order to recover the well-known $\kappa$-Minkowski spacetime~\cite{Maslanka1993} as well as the (3+1)D $\kappa$-(A)dS spacetimes~\cite{BGH2019kappaAdS3+1}. In Section~\ref{s4}, we present  other noncommutative  (3+1)D Minkowski and  (A)dS spacetimes,  which are quite different  from the usual $\kappa$-spacetimes ones, by requiring to preserve a  quantum Lorentz  subalgebra~\cite{Ballesteros:2021inq}.
 
 Now, we stress that in many  proposals to quantum gravity  theories from quantum groups   their cornerstone is usually  focused on the (3+1)D noncommutative spacetimes (in general, the $\kappa$-Minkowski spacetime), forgetting the role   that 6D quantum spaces of geodesics could be played. In fact, in our opinion,    any consistent theory should   consider, simultaneously,  both  a  (3+1)D  noncommutative spacetime and a 6D noncommutative space of worldlines. With this idea and by taking into account the very same  six-step procedure of Section~\ref{s2}, we construct the 6D $\kappa$-Poincar\'e quantum space  of time-like geodesics~\cite{BGH2019worldlinesplb} in Section~\ref{s5}.  Furthermore, there exist two other types of $\kappa$-Poincar\'e deformations  beyond the usual "time-like" one; namely, the  "space-like" and the  "light-like" deformations (see~\cite{BP2014extendedkappa,Geodesics2022} and references therein). Thus, we also present in Section~\ref{s5} these two remaining and   very recently obtained 6D noncommutative Poincar\'e  spaces of geodesics~\cite{Geodesics2022}.
 
  Finally, some remarks and open problems are addressed in the last section.


\section{Noncommutative spaces from   Poisson homogeneous spaces}
 \label{s2}

In this section, we firstly review  the   basic mathematical tools   necessary for the paper and, secondly, we  present   a general approach that allows one to construct noncommutative spaces from  coisotropic Poisson homogeneous spaces.

Let $G$ be a Lie group with Lie algebra  $\mathfrak{g}$ of dimension $d$. 
We consider a   Cartan decomposition of $\mathfrak{g}$, as a vector space,   given by the   sum of two subspaces
\be
{\mathfrak{g}}=  \mathfrak{h}\,{ \oplus }\,\mathfrak{t} , \qquad  [\mathfrak{h} ,\mathfrak {h} ] \subset \mathfrak{h}  .
\label{b1}
\ee
A generic  
  $\ell$-dimensional ($\ell$D)  homogeneous space   is defined as  the    left coset space  
\be
M^\ell=G/{H},
\label{b2}
\ee 
 where   $H$ is the $(d-\ell)$D isotropy subgroup  with   Lie algebra  $\mathfrak{h}$ (\ref{b1}). Hence we can identify the tangent space at every point $m = g H \in M^\ell$,  $g\in G$,  with the subspace $\mathfrak{t}$:
\begin{equation}
T_{m} (M^\ell) = T_{gH} (G/H) \simeq \mathfrak{g}/\mathfrak{h} \simeq \mathfrak{t} = \spn{ \{T_1,  \dots, T_\ell\}} .
\label{b3}
\end{equation}
The generators of the isotropy subalgebra $\mathfrak{h}$ keep a point on $M^\ell$ invariant,  the origin $O$,       playing the role of  rotations around $O$,  while the $\ell$ generators belonging to $\mathfrak{t}$ move $O$  along  $\ell$ basic directions,  thus behaving as translations on   $M^\ell$. The   local coordinates $(t^1,\dots,t^\ell)$  associated with the translation generators of $\mathfrak{t}$ (\ref{b3}) give rise  to  $\ell$ coordinates  on $M^\ell$.

A Poisson--Lie (PL) group is  a pair $(G,\Pi)$ where $G$ is a Lie group and $\Pi$ is a Poisson structure such that the Lie group multiplication $\mu : G \times G \rightarrow G$ is a Poisson map with respect to $\Pi$ on $G$ and  the product Poisson structure $\Pi_{G\times G} = \Pi \oplus \Pi$ on $G \times G$. The relation between the Poisson bivector field and the Poisson bracket is given by 
\begin{equation}
(\mathrm d f_1 \otimes \mathrm d f_2) \Pi = \{f_1,f_2 \}_\Pi  .
\label{b4}
\end{equation}

 A    Poisson manifold $(M,\pi)$  is  a manifold $M$ endowed  with a  Poisson structure $\pi$ on $M$. A Poisson homogeneous space (PHS) for a PL group $(G,\Pi)$ is a Poisson manifold $(M,\pi)$ which is endowed with a  transitive  group action   $\alpha : (G \times M, \Pi \oplus \pi ) \rightarrow (M, \pi)$  which is a Poisson map.  Throughout this paper we shall consider  that the manifold is a  homogeneous space $M\equiv M^\ell=G/H$ (\ref{b2}). In this case,  the Poisson structure $\pi$  on $M^\ell$ can be obtained by canonical projection of the PL structure $\Pi$ on $G$.
 
Next, a Lie bialgebra is a pair $(\mf g, \delta)$ where $\mf g$ is a Lie algebra and  $\delta : \mf g \to   \mf g\wedge \mf g$
is a linear map called the cocommutator   satisfying the following  two conditions~\cite{ChariPressley1994}:
 
 \noindent
 (i) $\delta$ is a 1-cocycle: 
\begin{equation}
\delta([X_i,X_j])=[\delta(X_i),\,  X_j\otimes 1+ 1\otimes X_j] + 
[ X_i\otimes 1+1\otimes X_i,\, \delta(X_j)] ,\quad\  \forall  X_i,X_j\in
\mathfrak{g}.
\label{b5} 
\end{equation}
 (ii) The dual map $\delta^\ast:\mathfrak{g}^\ast\otimes \mathfrak{g}^\ast \to \mathfrak{g}^\ast$ is a Lie bracket on the dual Lie algebra $\mathfrak{g}^\ast$ of $\mathfrak{g}$.

Coboundary Lie bialgebras~\cite{ChariPressley1994,Drinfeld1983hamiltonian} are those provided by a skewsymmetric  classical $r$-matrix   $r\in   \mathfrak{g} \wedge \mathfrak{g} $
in the form
\be
\delta(X_i)=[ X_i \otimes 1+1\otimes X_i ,\,  r],\qquad 
\forall X_i\in \mathfrak{g} ,
\label{b6}
\ee
such that $r$    must
  be a solution of the modified classical Yang--Baxter equation (mCYBE)  
\be
[X_i\otimes 1\otimes 1 + 1\otimes X_i\otimes 1 +
1\otimes 1\otimes X_i,[[r,r]]\, ]=0, \qquad \forall X_i\in \mathfrak{g},
\label{b7}
\ee
where $[[r,r]]$ is the  Schouten bracket    defined by
\be
[[r,r]]:=[r_{12},r_{13}]+ [r_{12},r_{23}]+ [r_{13},r_{23}] ,
\label{b8}
\ee
such that
  \be
  r_{12}= r^{ij} X_i \otimes X_j\otimes 1 , 
  \qquad
  r_{13}= r^{ij} X_i \otimes 1\otimes X_j  , 
  \qquad
  r_{23}= r^{ij} 1 \otimes X_i\otimes X_j  ,
\label{b9}
  \ee
and hereafter  sum over repeated indices will be understood  unless otherwise stated. If  the  Schouten bracket (\ref{b8}) does not vanish    the Lie algebra $\mathfrak{g}$ is  said to be endowed with a  quasitriangular or standard Lie bialgebra structure $(\mathfrak{g},\delta(r))$.    The vanishing of the Schouten bracket  (\ref{b8})  leads to the classical Yang--Baxter equation (CYBE)  $[[r,r]]=0$   and $(\mathfrak{g},\delta(r))$ is called a triangular or nonstandard  Lie bialgebra.

 The main point now is that    coboundary Lie bialgebras $(\mf g, \delta(r))$ are the tangent counterpart of coboundary PL groups $(G,\Pi)$ \cite{ChariPressley1994}, where the Poisson structure $\Pi$ on $G$ is given by the Sklyanin bracket
\begin{align}
\begin{split}
&\{f_1,f_2\}=r^{ij}\left( X^L_i f_1\, X^L_j f_2 - X^R_i f_1 \, X^R_j f_2 \right),\qquad f_1,f_2 \in \mathcal C (G),
\end{split}
\label{b10}
\end{align} 
such that    $X^L_i$ and $ X^R_i$ are   left- and right-invariant vector fields  defined by
\begin{align}
X^L_i f(g)&=\frac{\dd}{\dd t}\biggr\rvert _{t=0} f\left(g\, {\rm e}^{t Y_i}\right),  \qquad  X^R_i f(g)=\frac{\dd}{\dd t}\biggr\rvert _{t=0} f\left({\rm e}^{t Y_i} g\right),
\label{b11}
\end{align}
 where $f \in \mathcal C (G)$, $g \in G$ and $Y_i \in \mathfrak g$.   The quantization (as a Hopf algebra) of   
a PL  group $(G,\Pi)$ is just the corresponding quantum group.

Given a  PHS  $(M^\ell=G/H,\pi)$ with an underlying coboundary Lie bialgebra $(\mf g, \delta(r))$ of $(G,\Pi)$,    the Poisson structure $\pi$  on $M^\ell$, coming  from canonical projection of the PL structure $\Pi$ on $G$,
is only ensured to be well-defined whenever the     so-called    coisotropy condition  for the cocommutator $\delta$ with respect to the isotropy subalgebra $\mathfrak h$ of $H$ is fulfilled~\cite{Ciccoli2006, BGM2019coreductive,GH2021symmetry}, namely
 \be
\delta(\mathfrak h) \subset \mathfrak h \wedge \mathfrak g.
\label{b12}
\ee
 This    condition means that the commutation relations that define the noncommutative space $M_z^\ell$,  with underlying classical space  $M^\ell$ (\ref{b2}) and quantum deformation parameter $q={\rm e}^z$, at the first-order in all the quantum coordinates $(\hat t^1,\dots, \hat t^\ell)$ close on  a Lie subalgebra which is just the   annihilator $\mathfrak{h}^\perp$ of $\mathfrak{h}$ on the dual Lie algebra $\mathfrak g^*$:
 \be
\mathfrak{h}^\perp \equiv M^\ell_z .
\label{b13}
\ee
The duality between the generators of $\mathfrak{t}$  (\ref{b3}) and  the quantum coordinates  $(\hat t^1,\dots, \hat t^\ell)$ spanning $ M^\ell_z$  is determined by a canonical pairing given by the bilinear form
\be
\langle  \hat t^j,T_k \rangle=\delta_k^j  \,,\qquad \forall j,k.
\label{b13bis}
\ee
Noncommutative spaces  can finally be  obtained as    quantizations   of  coisotropic PHS  in all orders in the quantum   coordinates $(\hat t^1,\dots, \hat t^\ell)$, so completing the  initial  quantum space $ M^\ell_z$ (\ref{b13}) which just determines the Lie-algebraic (linear) contribution.

A general approach in order to construct  any noncommutative space   from any coisotropic PHS $(M^\ell=G/H,\pi)$ with coboundary Lie bialgebra $(\mf g, \delta(r))$, so fulfilling (\ref{b12}),   is   summarized in six steps (see \cite{
  Ballesteros:2021inq,Geodesics2022} and references therein) as follows:
\begin{enumerate}

\item Consider a faithful representation $\rho$ of the Lie  algebra $\mf g$.

\item Compute, by exponentiation, an element  of the Lie group $G$ according to the left coset space $M^\ell=G/{H}$ (\ref{b2})   in the form
\be
G_{M^\ell}= \exp{\!\bigl( t^1 \rho(T_1) \bigr)}  \cdots\, \exp\!{ \bigl( t^\ell \rho(T_\ell) \bigr)} \,H ,
\label{b14}
\ee
where $(T_1,  \dots, T_\ell)$ are the translation generators on $M^\ell$,  $H$ is the $(d-\ell)$D isotropy subgroup, and 
$(t^1,\dots,t^\ell)$ are   local coordinates   associated with the above translation generators of $\mathfrak{t}$ (\ref{b3}).

\item Calculate the corresponding left- and right-invariant vector fields (\ref{b11})  from $G_{M^\ell}$ (\ref{b14}).

\item Consider a classical $r$-matrix (\ref{b7}) so determining a coboundary Lie bialgebra  $(\mathfrak{g},\delta(r))$ (either  of quasitriangular or triangular type), which   is the tangent counterpart of  the  corresponding co\-boundary PL group $(G,\Pi)$.

\item
Obtain the Poisson brackets among the local translation       coordinates $(t^1,\dots,t^\ell)$ via the 
 Sklyanin bracket (\ref{b10}) from the chosen classical $r$-matrix. The resulting expressions define the  coisotropic PHS.

\item Finally, quantize the PHS thus obtaining the  noncommutative space in terms of the quantum coordinates 
$(\hat t^1,\dots,\hat t^\ell)$.

\end{enumerate}

 In the next sections we illustrate the above procedure by applying it to several    (A)dS and Poincar\'e quantum deformations giving rise to noncommutative  spaces that could  be relevant in a quantum gravity framework~\cite{Addazi:2021xuf}.


\section{$\kappa$-Minkowski and  $\kappa$-(A)dS   noncommutative spacetimes}
 \label{s3}

Let us consider the (3+1)D Poincar\'e and    (A)dS Lie algebras expressed as  a one-parametric family of Lie algebras  denoted by $\mathfrak g_\L$    depending     on the cosmological constant  $\Lambda$.  
 In a   kinematical basis spanned by the   generators of time translations $P_0$, spatial translations  $\>P=(P_1,P_2,P_3)$, boost transformations $\>K=(K_1,K_2,K_3)$  and rotations $\>J=(J_1,J_2,J_3)$, the      commutation relations of  $\mathfrak g_\L$ are given by
\begin{equation}
\begin{array}{lll}
[J_a,J_b]=\epsilon_{abc}J_c ,& \quad [J_a,P_b]=\epsilon_{abc}P_c , &\quad [J_a,K_b]=\epsilon_{abc}K_c , \\[2pt]
\displaystyle{
  [K_a,P_0]=P_a  } , &\quad\displaystyle{[K_a,P_b]=\delta_{ab} P_0} ,    &\quad
  \displaystyle{[K_a,K_b]=-\epsilon_{abc} J_c} , 
\\[2pt][P_0,P_a]=- \L  K_a , &\quad   [P_a,P_b]=\L \epsilon_{abc}J_c , &\quad[J_a,P_0]=0    .
\label{c1}

\end{array}
\end{equation}
  From now on, Latin indices run as $a,b,c=1,2,3$ while Greek ones run as  $\mu=0,1,2,3$. The  Lie algebra  $\mathfrak g_\L$ comprises the dS   algebra $\mathfrak{so}(4,1)$ for $\Lambda > 0$, the AdS   algebra $\mathfrak{so}(3,2)$ for $\Lambda < 0$ and the Poincar\'e one  $\mathfrak{iso}(3,1)$  when $\Lambda = 0$. 

The first step in our approach is to consider a   faithful representation  $\rho : \mathfrak g_\L \rightarrow \text{End}(\mathbb R ^5)$ for  $X\in \mathfrak g_\L$, that reads
\begin{equation}
\rho(X)=  x^\mu \rho(P_\mu)   +  \xi^a \rho(K_a) +  \theta^a \rho(J_a) =\left(\begin{array}{ccccc}
0&\L x^0&-\L x^1&-\L x^2&-\L x^3\cr 
x^0 &0&\xi^1&\xi^2&\xi^3\cr 
x^1 &\xi^1&0&-\theta^3&\theta^2\cr 
x^2 &\xi^2&\theta^3&0&-\theta^1\cr 
x^3 &\xi^3&-\theta^2&\theta^1&0
\end{array}\right) .
\label{c2}
\end{equation}
By exponentiation we obtain    a one-parametric family of Lie groups,     $G_\Lambda$, that covers   the dS     $\mathrm{SO}(4,1)$ for $\Lambda > 0$, the  AdS   $\mathrm{SO}(3,2)$  for $\Lambda < 0$, and  the Poincar\'e    $\mathrm{ISO}(3,1)$ for $\Lambda = 0$.   The  (3+1)D  Minkowski and (A)dS homogeneous spacetimes  (\ref{b2}),   $M^{3+1}_\L$, are defined by 
 \be
M^{3+1}_\L = G_\L/H,\qquad  H={\rm SO}(3,1)= \langle \>K,\>J \rangle,
\label{c3}
\ee
where the Lie algebra $\mf h$ of $H$ is the Lorentz subalgebra and $\mf t=\spn{ \{P_\mu\}} $ (\ref{b1}). Observe that the constant sectional curvature
of $M^{3+1}_\L $ is $\omega=-\L$.

 Our aim now is to construct the $\kappa$-noncommutative counterpart of $M^{3+1}_\L $ (\ref{c3}). According to  (\ref{b14})   (step 2 in Section~\ref{s2})
 we compute   $G_\Lambda$   in terms of  local coordinates $(x^\mu, \xi^a, \theta^a )$ as
 \begin{align}
G_{\L}= \exp{\!\bigl(x^0 \rho(P_0)\bigr)} \exp{\!\bigl(x^1 \rho(P_1)\bigr)} \exp{\!\bigl(x^2 \rho(P_2)\bigr)} \exp{\!\bigl(x^3 \rho(P_3)\bigr)} \, H,
\label{c4}
\end{align}
where the Lorentz subgroup $ H={\rm SO}(3,1)$ is parametrized by
\begin{align}
  H= \exp\bigl({\xi^1 \rho(K_1)}\bigr) \exp\bigl({\xi^2 \rho(K_2)}\bigr) \exp\bigl({\xi^3 \rho(K_3)}\bigr) \exp\bigl({\theta^1 \rho(J_1)}\bigr) \exp\bigl({\theta^2 \rho(J_2)} \bigr)\exp\bigl({\theta^3 \rho(J_3)}\bigr) .
  \label{c5}
\end{align}
Notice that here the index $\ell=4$ in   (\ref{b2}) and    the generic local coordinates    $(t^1,t^2,t^3,t^4)$ in  (\ref{b14})    corresponds to the spacetime coordinates $(x^0,x^1,x^2,x^3)$.

Following the step 3 in Section~\ref{s2}  we compute the    left- and right-invariant vector fields  (\ref{b11}) from $G_{\L}$. In the step 4 we have to consider a classical $r$-matrix and we distinguish two cases between $\kappa$-Poincar\'e  with $\L=0$ and $\kappa$-(A)dS with $\L \ne 0$.

The $\kappa$-Poincar\'e classical $r$-matrix is a solution of  the  mCYBE    (\ref{b7})
and reads~\cite{Maslanka1993,Zakrzewski1994poincare}
\be
r_0= \frac{1}{\kappa} \, ( K_1 \wedge P_1 + K_2 \wedge P_2 + K_3 \wedge P_3  )  ,
  \label{c6}
\ee
that satisfies the coisotropy condition (\ref{b12}) with respect to $\mf h=\spn\{ \>K,\>J\}$  and where the quantum deformation parameter $\kappa=1/z$.
 The corresponding  Sklyanin bracket (\ref{b10})  leads  to linear Poisson brackets for the classical coordinates $x^\mu$ which determine the $\kappa$-Minkowski PHS.  This can therefore be quantized directly in terms of the quantum coordinates   $\hat x^\mu$. Hence we recover well-known $\kappa$-Minkowski spacetime~\cite{Maslanka1993} (see also~\cite{GH2021symmetry,LukierskiRuegg1994,Daszkiewicz2008,BP2014extendedkappa} and references therein) which is of    Lie-algebraic type:
\be
[\hat x^0,\hat x^a]=-\frac 1{\kappa}\, \hat x^a,\qquad [\hat x^a,\hat x^b]=0,
\label{c7}
\ee
completing the final steps 5 and 6  in Section~\ref{s2}.

When  $\L \ne 0$ we consider the $\kappa$-(A)dS  classical $r$-matrix, which is also a  a solution of  the  mCYBE    (\ref{b7}), given by~\cite{BGHOS1995quasiorthogonal, BHMN2017kappa3+1,BGH2019kappaAdS3+1}
\be
r_\L=\frac{1}{\kappa}\,( K_1 \wedge P_1 + K_2 \wedge P_2 + K_3 \wedge P_3 + \ro  J_1 \wedge  J_2) ,
\label{c8}
\ee
such that the parameter $\ro$ is related to the cosmological constant $\L$ and the sectional curvature $\omega$ of the  (A)dS  spacetimes (\ref{c3}) by
\be
\omega = \ro^2= - \L.
\label{c9}
\ee
Thus $\ro$ is real for   AdS   and a purely imaginary number    for   dS.
The Sklyanin bracket  now gives rise   to the  (nonlinear) $\kappa$-(A)dS PHS in the form~\cite{BGH2019kappaAdS3+1}
\begin{align}
\begin{split}
&\{x^0,x^1\} =-\frac{1}{\kappa}\, \frac{\tanh (\ro x^1)}{\ro \cosh^2(\ro x^2) \cosh^2(\ro x^3)}\, ,\\
&\{x^0,x^2\} =-\frac{1}{\kappa}\,\frac{\tanh (\ro x^2)}{\ro \cosh^2(\ro x^3)} \,,\\
&\{x^0,x^3\} =-\frac{1}{\kappa}\,\frac{\tanh (\ro x^3)}{\ro}\, ,
\end{split}
\label{c10}
\end{align} 
\begin{align}
\begin{split}
&\{x^1,x^2\} =-\frac{1}{\kappa}\,\frac{\cosh (\ro x^1) \tanh ^2(\ro x^3)}{\ro}\, ,\\
&\{x^1,x^3\} =\frac{1}{\kappa}\,\frac{\cosh (\ro x^1) \tanh (\ro x^2) \tanh (\ro x^3)}{\ro}\, ,\\
&\{x^2,x^3\} =-\frac{1}{\kappa}\,\frac{\sinh (\ro x^1) \tanh (\ro x^3)}{\ro} \, .
\end{split}
\label{c11}
\end{align} 
Consequently, in contrast to the $\kappa$-Minkowski spacetime  (\ref{c7})  when $\L\ne 0$ the    3-space (\ref{c11}), determined by $  x^a$, is no longer commutative and ordering ambiguities arise in (\ref{c10}) and (\ref{c11}) which precludes a direct quantization.  This problem can be circumvented by introducing five ambient coordinates in the (A)dS spacetimes (\ref{c3}) denoted $(s^4,s^\mu)\in \mathbb R^{5}$ such that  they fulfil the pseudosphere constraint
\be
\Sigma_\L\equiv( s^4)^2 - \L  (s^0)^2 +\L \bigl( (s^1)^2+ (s^2)^2+ (s^3)^2 \bigr)=1 .
\label{c12}
\ee
These read\cite{Ballesteros:2021inq,BGH2019kappaAdS3+1}
 \begin{align} 
\begin{split}
&s^4=\cos( \ro x^0) \cosh( \ro x^1) \cosh( \ro x^2 )\cosh( \ro x^3 ), \\
&s^0=\frac {\sin( \ro x^0)}\ro  \cosh( \ro x^1 )\cosh( \ro x^2 )\cosh( \ro x^3), \\
&s^1=\frac {\sinh( \ro x^1) }\ro   \cosh (\ro x^2) \cosh( \ro x^3), \\
&s^2=\frac { \sinh (\ro x^2)} \ro\cosh( \ro x^3), \\
&s^3=\frac { \sinh (\ro x^3)} \ro ,
\end{split}
\label{c13}
\end{align}
and the  spacetime coordinates $x^\mu$ are  called geodesic
parallel coordinates. Notice also that  $
q^\mu= {s^\mu}/{s^4}$ are      Beltrami projective coordinates in $M^{3+1}_\L$  (\ref{c3})  which can be obtained through
 the  projection with pole  
$(0,0,0,0,0)\in \mathbb R^{5}$ of a point with ambient coordinates $(s^4,s^\mu) $ onto the projective hyperplane with $s^4=+1$ (see~\cite{BGH2020snyder} for details).    Next, if we compute the Poisson brackets among $(s^4,s^\mu)$ from (\ref{c10}) and (\ref{c11}), consider the quantum coordinates   $(\hat s^4, \hat s^\mu)$ along with   the  ordered monomials $(\hat s^0)^k\,(\hat s^1)^l\,(\hat s^3)^m\,(\hat s^2)^n\,(\hat s^4)^j$, we finally obtain  the    $\kappa$-(A)dS   spacetimes $M^{3+1}_{\L,\kappa}$ expressed as a quadratic algebra~\cite{BGH2019kappaAdS3+1}

 \begin{equation}
\begin{aligned}
&  [\hat s^0, \hat s^a] = -\frac{1}{\kappa}\, \hat s^{a} \hat s^{4} ,\qquad [\hat s^4,\hat s^a] = \frac{\ro^{2}}{\kappa}\, \hat s^{0}  \hat s^{a}  ,\qquad [\hat s^0,\hat s^4] = -\frac{\ro^{2}}{\kappa} \, \hat \esf_{\ro/\kappa} ,\\[2pt]
 &[\hat s^1,\hat s^2] = -\frac{\ro}{\kappa} (\hat s^{3})^{2} ,\qquad
 [\hat s^1,\hat s^3] = \frac{\ro}{\kappa}\, \hat s^{3}  \hat s^{2} ,\qquad
 [\hat s^2,\hat s^3] = -\frac{\ro}{\kappa}\, \hat s^{1} \hat s^{3} ,
 \label{c14}
 \end{aligned}
\end{equation}
where the quantum 3-space $\hat \esf_{\ro/\kappa} $ operator is given by
\be
\hat \esf_{\ro/\kappa}= (\hat s^1)^2 + (\hat s^2)^2 + (\hat s^3)^2  + \frac{\ro}{\kappa}\, \hat s^1 \hat s^2  .
\label{c15}
\ee
Obviously, Jacobi identities  are satisfied.
We remark that $M^{3+1}_{\L,\kappa}$  (\ref{c14}) has a   Casimir operator  
\be
\hat \Sigma_{\L,\kappa}= (\hat s^4)^2 -\L (\hat s^0)^2 + \frac{\L}{\kappa} \, \hat s^0 \hat s^4 +\L \,\hat \esf_{\ro/\kappa} ,
\label{c116}
\ee
which is   the quantum analogue of the pseudosphere (\ref{c12}) (recall that $\L=-\ro^2$ (\ref{c9})).

As expected,     under the flat limit  $\ro\to 0$ (i.e., $\L\to 0$), the ambient coordinates $(s^4,s^\mu)$ (\ref{c13}) provide   the usual Cartesian ones 
$(1,x^\mu)$ in the Minkowski spacetime and the  $\kappa$-(A)dS   spacetimes (\ref{c14})  reduce to the 
 $\kappa$-Minkowski   spacetime (\ref{c7}).


\section{Noncommutative (A)dS and Minkowski  spacetimes with quantum Lorentz subgroups}
\label{s4}

In this section we present   very recent results concerning (3+1)D noncommutative (A)dS and Minkowski  spacetimes that preserve a quantum Lorentz subgroup which were   obtained in~\cite{Ballesteros:2021inq} by following the same six-step procedure described in Section~\ref{s2}. We advance that these   are quite different from the   $\kappa$-Minkowski     (\ref{c7})  and $\kappa$-(A)dS   (\ref{c14})    spacetimes reviewed in the previous section. Hence, we keep the same notation as in    Section~\ref{s3}, in particular we shall make use of the expressions (\ref{c1})--(\ref{c5}), (\ref{c9}), (\ref{c12}) and (\ref{c13}).

We consider the family of the   (3+1)D  Poincar\'e and (A)dS   Lie algebras $\mathfrak g_\L$  (\ref{c1}) and search for classical $r$-matrices (\ref{b7})  that keep the Lorentz subalgebra $\mf h= \spn\{\>K,\>J  \}=\mf{so}(3,1)$ as a  sub-Lie bialgebra, that is, 
\be
\delta\left(\mathfrak{h}\right) \subset \mathfrak{h} \wedge \mathfrak{h},
\label{d1}
\ee
which is a  more restrictive version of the coisotropy condition (\ref{b12}). 
This restriction implies that  the corresponding PHS  is constructed through the Lorentz isotropy subgroup $ H={\rm SO}(3,1)$  such that   $(H,\Pi |_H)$  is a PL subgroup of  $(G_\L,\Pi)$ and it
 is called a PHS of Poisson subgroup type. 

 Then we start with the most general  element $r \in  \mf g_\L\wedge \mf g_\L$. Since the dimension of $\mf g_\L$ is $d=10$, $r$    depends on  45 initial deformation parameters. From it, we   directly compute the cocommutator   $\delta$ (\ref{b6}) such that $(\mf g_\L, \delta(r))$ defines a Lie bialgebra   if and only if $r$ is a solution of the mCYBE (\ref{b7}).
Moreover, we have to impose the condition (\ref{d1}).  

The simplest case is to require that  $\delta\left(\mathfrak{h}\right)=0$ which  means that the Lorentz subgroup remains underformed. The final result is summarized as~\cite{Ballesteros:2021inq}:
 \begin{proposition}
 \label{prop1} 
The only PL group $(G_\L, \Pi)$ such that $\Pi |_H = 0$ is the trivial one. 
\end{proposition}
Therefore 
the only PHS  $(M^{3+1}_\L=G_\L/H, \pi )$ of Poisson Lorentz subgroup type such that $\Pi |_H = 0$ is the trivial one. 
In other words, there does not exist any quantum deformation of the  (3+1)D Poincar\'e and (A)dS Lie algebras preserving 
the Lorentz subalgebra $\mf h$ underformed.

Now the main question is whether there exists a quantum deformation of $\mf g_\L$ preserving a non-trivial quantum Lorentz subalgebra, that is, $\delta\left(\mathfrak{h}\right) \subset \mathfrak{h} \wedge \mathfrak{h}\ne 0$. The answer is positive. By taking into account previous results concerning quantum Poincar\'e groups~\cite{PW1996,Zakrzewski1997} and  quantum  deformations of the Lorentz algebra $\mf h=\mf{so}(3,1)$~\cite{Zakrzewski1994lorentz}, it can be proven that the classification of  the quantum deformations of $\mf g_\L$ keeping a quantum Lorentz subalgebra can be casted into three types as follows~\cite{Ballesteros:2021inq}:

\begin{proposition}
\label{prop2}
There exist three classes of PHS $(M^{3+1}_\L=G_\L/H, \pi )$ for  each of the maximally symmetric relativistic spacetimes of constant curvature (Minkowski and (A)dS) (\ref{c3}) such that the isotropy Lorentz subgroup  $H$ is a PL subgroup of $(G_\L, \Pi)$. All of them are obtained from coboundary PL  structures on their respective isometry group $G_\L$ which are determined, up to  $\mf g_\L$-isomorphisms,  by the classical  $r$-matrices
\begin{equation}
\begin{split}
r_{\rm I} = \, &z\, (K_1 \w K_2 + K_1 \w J_3 - K_3 \w J_1 - J_1 \w J_2) \\
&\quad \ - z' \,(K_2 \w K_3 - K_2 \w J_2 - K_3 \w J_3 + J_2 \w J_3) ,\\
r_{\rm II} = \, &z \,K_1 \w J_1 ,\\
r_{\rm III} =\,  &z \,(K_1 \w K_2 + K_1 \w J_3) ,\\
\end{split}
\label{d2}
\end{equation}
where $z$ and $z'$ are free quantum deformation parameters. These three classical $r$-matrices are solutions of the CYBE  $[[r,r]]=0$.
\end{proposition}
Hence the three classes correspond to triangular or nonstandard deformations. Types II and III would provide one-parametric deformations, while   type I would lead to a two-parametric one with arbitrary deformation parameters $z$ and $z'$. Recall that the $\kappa$-$\mf g_\L$ deformations  described in the previous section have a quasitriangular or standard character.

Next we apply the approach presented in Section~\ref{s2} in order to construct the corresponding PHS from the above classical $r$-matrices in terms of the local coordinates $x^\mu$ through the Sklyanin bracket (\ref{b10}). However, the resulting expressions are rather cumbersome and strong ordering ambiguities appear, so there is no a direct quantization for any class. In order to solve this problem we proceed similarly  to the  $\kappa$-(A)dS 
spacetimes (\ref{c14}). We  again consider  the  ambient coordinates $(s^4,s^\mu) $ (\ref{c13}) (subjected to    the pseudosphere  constraint (\ref{c12})),  compute their PL brackets from those initially given in terms of $x^\mu$, and finally obtain the corresponding noncommutive spacetimes by choosing an appropriate order in the quantum coordinates 
$(\hat s^4,\hat s^\mu) $ (so satisfying the Jacobi identities).

As a final result, we display   in Table~\ref{table1} all the  (3+1)D Minkowski and (A)dS noncommutive spacetimes that preserve a non-trivial quantum Lorentz subgroup~\cite{Ballesteros:2021inq}.

  
\begin{table}[t]
{\small
\caption{\small  The three types of (A)dS and Minkowski  noncommutive   spacetimes with  quantum Lorentz subgroups    determined by Proposition~\ref{prop2}.  These are expressed
 in    quantum ambient spacetime coordinates  $\hat s^\mu$ (\ref{c13})  or in $(\hat s^\pm=\hat s^0\pm \hat s^1,\hat s^2,\hat s^3)$. The quantum coordinate $\hat s^4$ always  commutes with $\hat s^\mu$. 
}
\label{table1}
  \begin{center}
\begin{tabular}{l  }
\hline

\hline
\\[-0.2cm]
 {{\bf Type I}\quad $r_{\rm I}=  z(K_1\wedge K_2+K_1\wedge J_3-K_3\wedge J_1-J_1\wedge J_2 ) $}  \\[0.2cm]
\qquad \qquad\qquad\quad $ - z'(K_2\wedge K_3-K_2\wedge J_2-K_3\wedge J_3+J_2\wedge J_3 )$  \\[0.2cm]
$\bullet$ Subfamily with $z=0$ \\[0.2cm]
 $ [\hat s^-, \hat s^2]= -2 z' \hat s^+ \hat s^3   \qquad
[\hat s^-,\hat s^3]=  2z' \hat s^+ \hat s^2 \qquad [\hat s^2,\hat s^3]= z'   (\hat s^+)^2  \qquad
[\hat s^+, \, \cdot\,]=0 $   \\[0.3cm]  
$\bullet$ Subfamily with $z'=0$ \\[0.2cm]
$  [\hat s^-, \hat s^+]=2 z   \hat s^+\hat s^2 \qquad  [\hat s^-,\hat s^2]=z\hat s^-  \hat s^+  - 2 z  ( \hat s^3)^2 \qquad 
 [\hat s^-, \hat s^3]=2 z   \hat s^3\hat s^2 $  \\[4pt]
$  [\hat s^2, \hat s^3]=z  \hat s^+\hat s^3 \qquad \ \ \  [\hat s^+, \hat s^2]=- z   (\hat s^+)^2 \qquad\qquad\quad\,   [\hat s^+, \hat s^3]=0  $   \\[0.25cm]
  \hline
\\[-0.2cm]
 {{\bf Type II}\quad  $r_{\rm II}=  zK_1\wedge J_1$} \\[0.2cm]
$[ \hat s^0, \hat s^1 ]=0  \qquad  [\hat s^0, \hat s^2] =z \hat s^1 \hat s^3 \qquad
[ \hat s^0,\hat s^3] =-z\hat  s^1 \hat s^2 $    \\[4pt]
$[ \hat s^2,\hat s^3] =0 \qquad [ \hat s^1,\hat s^2] =z\hat s^0 \hat s^3 \qquad  [\hat s^1,\hat s^3]=-z\hat s^0\hat s^2    $ 
 \\[0.25cm]
 \hline
\\[-0.2cm]
 {{\bf Type III}\quad  $r_{\rm III}=  z(K_1\wedge K_2+K_1\wedge J_3 )$}  \\[0.2cm]
$ [\hat s^2, \hat s^+]=z (\hat s^+)^2  \qquad  [\hat s^2,\hat s^-]=-z \hat s^- \hat s^+    \qquad
[\hat  s^-,\hat s^+]=2 z \hat s^+  \hat  s^2       \qquad  [\hat  s^3,\hat s^\mu]=0    $ 
 \\[0.3cm]
\hline

\hline
\\[-0.2cm]
\end{tabular}
 \end{center}
}
 \end{table} 
 

Now some remarks are in order.
\begin{itemize}

\item The ambient quantum  coordinate $\hat s^4$  is always a central element for all  the  three types of  noncommutive spacetimes, 
 $[   \hat s^4 ,\hat s^\mu]=0$,  so that these are just defined by the (3+1) quantum variables $\hat s^\mu$.

\item In this respect, we remark that the corresponding  noncommutive Minkowski spacetimes can directly be obtained through
 the flat limit $\L \to 0$ (or $\eta\to 0)$, in such a manner that the quantum coordinates 
$(\hat s^4,\hat s^\mu) $ reduce to the usual quantum Cartesian ones $(  1,\hat x^\mu) $. Since $\hat s^4$ is absent in all the expressions presented in Table~\ref{table1}, the noncommutive Minkowski spacetimes  adopt the very same formal expressions in the quantum  Cartesian coordinates $\hat x^\mu$.

\item For types I and III it is found that the explicit noncommutive spacetimes are naturally adapted to a null-plane basis~\cite{Ballesteros:2021inq,Leutwyler78} and for this reason we have considered the quantum coordinates $(\hat s^\pm=\hat s^0\pm \hat s^1,\hat s^2,\hat s^3)$ instead of $\hat s^\mu$. Thus they lead to   
 $(\hat x^\pm=\hat x^0\pm \hat x^1,\hat x^2,\hat x^3)$ for the Minkowski cases.

\item In type I we have distinguished two subfamilies with either $z$ or $z'$ equal to zero in order to clarify the presentation of the results. Nevertheless, observe that the general noncommutive spacetimes of type I is just the superposition (the sum) of both subfamilies.

\item We remark that the type II noncommutative spacetime has already been  obtained for the quadratic Minkowski case in~\cite{Lukierski2006}   (set $\hat s^\mu\equiv \hat x^\mu$) by following a different approach from ours;  that is, from  a twisted quantum Poincar\'e group and then applying the FRT procedure. Notice that, in fact,  the classical $r$-matrix $r_{\rm II}$ (\ref{d2}) is just a Reshetikhin twist.

 \item Finally,   the type III noncommutative spacetimes can be regarded as   (2+1)D quantum spaces since  the quantum coordinate $\hat s^3$ is a central operator,  
 $ [\hat  s^3, \, \cdot\,]=0$. We recall that when this structure   is, again,  only applied to the Minkowski case $(\hat x^\pm=\hat x^0\pm \hat x^1,\hat x^2)$,  it was  already obtained from a Drinfel'd double structure  of the (2+1)D  Poincar\'e group in~\cite{PoinDD}. In addition,  we stress that the corresponding quantum algebra for $\mf g_\L$  comes  from   the lower dimensional  Lorentz subalgebra $\mf {so}(2,1)$ spanned by $\{J_3,K_1,K_2\}$    which is just  the well-known nonstandard (or Jordanian) quantum deformation of $\mf {sl}(2,\mathbb R)\simeq \mf {so}(2,1)$ (see~\cite{Ohn,Ogievetsky,BH1996,Shariati}).  
 For  higher-dimensional quantum (A)dS  algebras  keeping such a  nonstandard quantum  $\mf {sl}(2,\mathbb R)$ Hopf subalgebra  we refer to~\cite{Herranz2002}.
 
\end{itemize}


\section{$\kappa$-Poincar\'e space of time-like worldlines and beyond}
\label{s5}

So far we have constructed several   (3+1)D Minkowski and (A)dS noncommutive spacetimes by applying the approach given in Section~\ref{s2}. However, we stress that such a procedure is rather general and can be applied to any homogeneous space. Hence in this section we shall consider the 6D homogeneous space of time-like Poincar\'e  geodesics and obtain its  $\kappa$-noncommutative version~\cite{BGH2019worldlinesplb}.

With this aim we consider the following  Cartan decomposition of the Poincar\'e algebra $\mf g_\L \equiv\mf g$  and $G_\L\equiv G$ with commutation relations (\ref{c1}) with $\L=0$  (see (\ref{b1})):
\be
\mathfrak g= \mathfrak t_\til \oplus \mathfrak h_\til ,  \qquad  \mathfrak{t}_\til = \spn \{  \>{P},  \> {K} \}    ,\qquad     \mathfrak h_\til = \spn\{P_0,  \> {J} \}=\mathbb R\oplus \mathfrak{so}(3).
\label{f1}
\ee
The  homogeneous space   of time-like geodesics is of dimension six and is defined by
\be 
\mathcal{W}_\til  =G/H_\til  ,
\label{f2}
\ee
where the isotropy subgroup $H_\til = \mathbb R\otimes {\rm SO}(3)$ comes from   the Lie subalgebra $\mathfrak h_\til $ (\ref{f1}).

By following the procedure presented in Section~\ref{s2}, 
 we first parametrize the Poincar\'e Lie group from the 5D matrix representation (\ref{c2}) with $\L=0$ taking into account the order given in (\ref{b14}), that is,
\begin{align}
\begin{split}
&G_{\mathcal{W}_{\til} } = \exp\bigl({\eta^1 \rho(K_1)}\bigr) \exp\bigl({y^1 \rho(P_1)}\bigr) \exp\bigl({\eta^2 \rho(K_2)}\bigr) \exp\bigl({y^2 \rho(P_2)} \bigr) \\
& \qquad\qquad\qquad   \times \exp\bigl({\eta^3 \rho(K_3)}\bigr) \exp\bigl({y^3 \rho(P_3)}\bigr) \,H_{\til} ,
\end{split}
\label{f3}
\end{align}
where $H_{\til}$ is    the stabilizer  of the worldline corresponding to a massive particle at rest at the origin of the (3+1)D Minkowski spacetime, namely
\begin{align}
\begin{split}
&H_{\til} = \exp\bigl( {\phi^1 \rho(J_1)}\bigr) \exp\bigl( {\phi^2 \rho(J_2)}\bigr) \exp\bigl( {\phi^3 \rho(J_3)}\bigr) \exp\bigl( {y^0 \rho(P_0)} \bigr).
\end{split}
\label{f4}
\end{align}
Therefore the classical coordinates $  
  (t^1,\dots,t^\ell)$ in (\ref{b14}) correspond to $({\eta^a,y^a})$ in (\ref{f3}) (recall that now $\ell=6$).
 Next we consider the $\kappa$-Poincar\'e   $r$-matrix   (\ref{c6})   and by projecting the Sklyanin bracket  (\ref{b10}) to the homogeneous space  ({\ref{f2})   we obtain a coisotropic PHS  for the classical space of time-like geodesics which can be straightforwardly quantized since no ordering problems appear.  In this way, the  $\kappa$-Poincar\'e space  of time-like geodesics $\mathcal{W}_{\til,\kappa}$  in terms of the six quantum coordinates $(\hat y^a,\hat \eta^a)$ turns out to be~\cite{BGH2019worldlinesplb}:
\begin{equation}
\begin{split}
[\hat y^1, \hat y^2] &= \frac{1}{\kappa} \left( \hat y^2 \sinh \hat \eta^1-\frac{\hat y^1 \tanh \hat \eta^2}{\cosh \hat \eta^3}\right) ,\\
[\hat y^1,\hat y^3] &= \frac{1}{\kappa} \big(\hat y^3 \sinh  \hat \eta^1 -\hat y^1 \tanh  \hat \eta^3 \big), \\
[\hat y^2, \hat y^3] &= \frac{1}{\kappa} \big(\hat y^3 \cosh  \hat \eta^1  \sinh  \hat \eta^2 -\hat y^2 \tanh  \hat \eta^3 \big), \\
[\hat y^1, \hat \eta^1] &= \frac{1}{\kappa} \,\frac{ \bigl(\cosh  \hat \eta^1  \cosh  \hat \eta^2  \cosh  \hat \eta^3 -1\bigr)}{\cosh  \hat \eta^2  \cosh  \hat \eta^3 }, \\
[\hat y^2, \hat \eta^2] &= \frac{1}{\kappa}\, \frac{ \bigl(\cosh  \hat \eta^1  \cosh  \hat \eta^2  \cosh  \hat \eta^3 -1\bigr)}{\cosh  \hat \eta^3 }, \\
[\hat y^3, \hat \eta^3] &= \frac{1}{\kappa}\, \bigl(\cosh  \hat \eta^1  \cosh  \hat \eta^2  \cosh  \hat \eta^3 -1\bigr) \, ,
\label{f5}
\end{split}
\end{equation}
together with
\be
[\hat \eta^a, \hat\eta^b] = 0,\quad  \forall a,b,\qquad [\hat y^a,\hat \eta^b] = 0,\quad    a \neq b. 
\label{f6}
\ee
The above commutators can also be written in terms of quantum Darboux operators
 $(\hat q^a,\hat p^a)$ on a 6D smooth submanifold   $(\eta^1,\eta^2,\eta^3) \neq (0,0,0)$; these are defined by
\begin{equation}
\begin{split}
 \hat q^1&:= \frac{ \cosh   \hat  \eta^2  \cosh \hat  \eta^3 }{\cosh   \hat  \eta^1  \cosh   \hat  \eta^2  \cosh   \hat  \eta^3 -1} \,  \hat  y^1 , \\
  \hat  q^2&:= \frac{ \cosh   \hat  \eta^3 }{\cosh   \hat  \eta^1  \cosh \hat    \eta^2  \cosh \hat   \eta^3 -1} \,   \hat  y^2 , \\
  \hat  q^3&:= \frac{ 1}{\cosh   \hat  \eta^1  \cosh   \hat  \eta^2  \cosh    \hat  \eta^3 -1} \,  \hat  y^3, \\
 \hat   p^a &:=  \hat  \eta^a ,
\label{f7}
\end{split}
\end{equation}
where the ordering $(\hat \eta^a)^m\,(\hat y^a)^n$ has to be preserved. They   lead to   the canonical commutation relations
\begin{align}
\big[ \hat q^a,\hat q^b\big]= \big[\hat p^a,\hat p^b\big]= 0, \qquad   \big[ \hat q^a,\hat p^b\big]= \frac 1 {\kappa} \, \delta_{ab}  .
\label{f8}
\end{align}
From these   expressions we find that  the noncommutative space  $\mathcal{W}_{\til,\kappa}$ can be regarded as  three copies of the usual Heisenberg--Weyl  algebra   of quantum mechanics where the deformation parameter $\kappa^{-1}$ replaces the Planck constant $\hbar$.  We also recall that  a first phenomenological analysis for  $\mathcal{W}_{\til,\kappa}$, expressed in the form (\ref{f5}) and (\ref{f6}),  was    performed in~\cite{BGGM2021fuzzy}.

So far we have constructed the noncommutative space  $\mathcal{W}_{\til,\kappa}$ from the usual "time-like" $\kappa$-Poincar\'e deformation with classical $r$-matrix (\ref{c6}). However we remark that there exist two other possible 
$\kappa$-Poincar\'e deformations provided by  "space-like" and "light-like"  classical $r$-matrices~\cite{BP2014extendedkappa,BGH2019worldlinesplb}. The quantization procedure described in Section~\ref{s2} can similarly be applied to these remaining cases in order to construct the quantum counterpart of the  6D homogeneous
space $\mathcal{W}_\til $ (\ref{f2}). Therefore we shall keep exactly the expressions (\ref{f3}) and (\ref{f4}) together with the associated invariant vector fields and only change the underlying $r$-matrix. In what follows we summarize the final results which were recently obtained in~\cite{Geodesics2022}.

We consider the  "space-like" $r$-matrix given by
\be
r =  \frac 1\kappa\left(K_3\wedge P_0 +J_1\wedge P_2 -J_2\wedge P_1 \right) ,
\label{f9}
\ee
which is also   a solution of the mCYBE (\ref{b7}), so quasitriangular. The corresponding quantum Poincar\'e algebra was obtained in~\cite{CK4d} (c.f.~Type 1.~(a) with  $z=1/\kappa$). When computing the PHS it is found that again there are no ordering problems so that this can be quantized directly leading to the commutation relations defining
$\mathcal{W}_{\til,\kappa}$  from the  "space-like" $\kappa$-Poincar\'e deformation; these are 
\begin{equation}
\begin{split}
[\hat y^1,\hat y^2]  &= - \frac{1}{\kappa}\, \hat y^1 \tanh  \hat \eta^2  \tanh  \hat \eta^3,  \\
[\hat y^1, \hat y^3]  &= \frac{1}{\kappa}\, \frac{\hat y^1}{\cosh  \hat \eta^3 } ,\\
[\hat y^2, \hat y^3]  &= \frac{1}{\kappa} \,\frac{\hat y^2}{\cosh  \hat \eta^3 }, \\
[\hat y^1, \hat \eta^1]  &= -\frac{1}{\kappa}\, \frac{\tanh  \hat \eta^3 }{\cosh  \hat \eta^2 } ,\\
[\hat y^2, \hat \eta^2]  &= -\frac{1}{\kappa} \tanh  \hat \eta^3  ,\\
[\hat y^3, \hat \eta^3]  &= -\frac{1}{\kappa} \sinh  \hat \eta^3 ,
\label{f10}
\end{split}
\end{equation}
  with the same vanishing brackets given by (\ref{f6}).

Finally, in the kinematical basis (\ref{c1}) with $\L=0$ the  "light-like" $\kappa$-Poincar\'e deformation is determined by  
\be
r =  \frac 1\kappa\left(K_3\wedge P_0 +K_1\wedge P_1 +K_2\wedge P_2 + K_3\wedge P_3 +J_1\wedge P_2 -J_2\wedge P_1 \right)  ,
\label{f11}
\ee
which is triangular with vanishing Schouten bracket. This element provides    the  so-called "null-plane" quantum Poincar\'e algebra introduced in~\cite{nullplane95,Rnullplane97} (where $z=1/\kappa$) in terms of a  null-plane basis~\cite{Leutwyler78} instead of the kinematical one. Notice that the  "light-like" $r$-matrix (\ref{f11}) is just the sum of the "time-like"  $r$-matrix   (\ref{c6})  and the  "space-like"  one (\ref{f9}). Consequently, as expected,   
the resulting PHS can directly be  quantized giving rise to $\mathcal{W}_{\til,\kappa}$  from the  "light-like" $\kappa$-Poincar\'e deformation which turns out to be given by the sum of (\ref{f5}) and (\ref{f10}) (preserving the same vanishing brackets (\ref{f6})); namely
\begin{equation}
\begin{split}
[\hat y^1,\hat y^2]  &= \frac{1}{\kappa} \left(\hat y^2 \sinh  \hat \eta^1 -\frac{\hat y^1  \tanh \hat \eta^2  \bigl( \sinh  \hat \eta^3 +1\bigr)}{\cosh  \hat \eta^3 }\right) ,\\
[\hat y^1,\hat y^3]  &= \frac{1}{\kappa} \left( \hat y^3 \sinh  \hat \eta^1 -\frac{\hat y^1 \bigl( \sinh  \hat \eta^3 -1\bigr)}{\cosh  \hat \eta^3 }\right) ,\\
[\hat y^2, \hat y^3]  &= \frac{1}{\kappa} \left(\hat y^3 \cosh  \hat \eta^1  \sinh  \hat \eta^2 -\frac{\hat y^2 (\sinh  \hat \eta^3 -1)}{\cosh  \hat \eta^3 }\right),\\
[\hat y^1, \hat \eta^1]  &= \frac{1}{\kappa} \left( \frac{ \cosh  \hat \eta^1  \cosh  \hat \eta^2  \cosh  \hat \eta^3-\sinh  \hat \eta^3 -1}{\cosh  \hat \eta^2  \cosh  \hat \eta^3 }  \right), \\
[\hat y^2, \hat \eta^2]  &= \frac{1}{\kappa} \left( \frac{\cosh  \hat \eta^1  \cosh  \hat \eta^2  \cosh  \hat \eta^3 - \sinh  \hat \eta^3 -1}{\cosh  \hat \eta^3 } \right) ,\\
[\hat y^3, \hat \eta^3]  &= \frac{1}{\kappa} \left( \cosh  \hat \eta^1  \cosh  \hat \eta^2  \cosh  \hat \eta^3- \sinh  \hat \eta^3   -1   \right) . 
\label{eq:comm_t_rl2}
\end{split}
\end{equation}
We remark that quantum Darboux operators
 $(\hat q^a,\hat p^a)$ satisfying (\ref{f8})  can also be  defined for these latter noncommutative spaces~\cite{Geodesics2022}.


\section{Concluding remarks and open problems}

In this "twofold" contribution we have, firstly, presented in Section~\ref{s2}  a general approach  to construct   noncommutative spaces from coisotropic PHS spaces determined by a coboundary Lie bialgebra structure  and, secondly,  we  have applied  it  to the physically relevant  (3+1)D (A)dS and Poincar\'e Lie groups.
Besides the well-known  (3+1)D $\kappa$-spacetimes  shown in Section~\ref{s3},  we have also presented  quite different (i.e.~non-equivalent)  (3+1)D noncommutative (A)dS  and Minkowski spacetimes by requiring to preserve a quantum Lorentz subgroup invariant  in Section~\ref{s4}. In addition, we have also considered  noncommutative spaces beyond  the   (3+1)D noncommutative spacetimes, which are the usual models considered in quantum gravity. In this respect, we have presented the only three possible 6D noncommutative spaces of time-like geodesics  provided the three types of $
\kappa$-Poincar\'e quantum  deformations in Section~\ref{s5}. We stress that a classification of all  6D noncommutative spaces of $
\kappa$-Poincar\'e    geodesics, covering the usual time-like deformation, already here described,  along with the space-like and light-like deformations can  be found in~\cite{Geodesics2022}.

To conclude, we would like to comment on some open problems. Obviously, the procedure  considered here can be applied to any coisotropic PHS space providing new  noncommutative spaces.  As far as (3+1)D (A)dS and Minkowski  noncommutative spacetimes are concerned, we have presented 
their well-known $\kappa$-deformation together with all  possible quantum spacetimes preserving a non-trivial quantum Lorentz subgroup. These results constitute the cornerstone of a  large number of possibilities for a further development.  Nevertheless, we remark that quantum spaces of geodesics have not been considered and studied so deeply. In fact, to the best of our knowledge,  only $\kappa$-deformations for quantum  Poincar\'e geodesics
have been achieved. This fact   not only suggests the consideration of other types of quantum  Poincar\'e geodesics but, in our opinion,  the relevant open problem is   to construct quantum (A)dS spaces of geodescics; there are no results on this problem from a quantum group setting. In fact, for the $\kappa$-Poincar\'e space of time-like worldlines (from the usual  $\kappa$-Poincar\'e  algebra) its  fuzzy properties  have been studied in~\cite{BGGM2021fuzzy} and by following~\cite{LMMP2018localization,BGGM2021fuzzy} a similar analysis could be faced with the other types of  $\kappa$-Poincar\'e geodesics. Consequently, the construction of  (A)dS    noncommutative  spaces of geodesics  (covariant under their corresponding  (A)dS quantum groups)  could be achieved   following the   same approach here presented, and thus the role of a nonvanishing cosmological constant  in this novel noncommutative geometric setting could be further analysed. Work on all these lines is in progress.


\section*{Acknowledgements}

 This work has been partially supported by Agencia Estatal de Investigaci\'on (Spain)  under grant  PID2019-106802GB-I00/AEI/10.13039/501100011033, by the Regional Government of Castilla y Le\'on (Junta de Castilla y Le\'on, Spain) and by the Spanish Ministry of Science and Innovation MICIN and the European Union NextGenerationEU/PRTR. The authors would like to acknowledge the contribution of the European Cooperation in Science and Technology COST Action CA18108.


\end{document}